\documentclass{article}
\usepackage{graphicx}

\begin{document}

\title{Multifragmentation, Clustering, and Coalescence in Nuclear Collisions}

\author{Stefan Scherer and Horst St\"ocker}

\date{\small\it Institut f\"ur Theoretische Physik,\\
Johann Wolfgang Goethe Universit\"at,\\
Robert Mayer-Str.\ 10,\\
D-60054 Frankfurt am Main, Germany \\ 
E-mail: scherer@th.physik.uni-frankfurt.de,\\
stoecker@uni-frankfurt.de}

\maketitle

\begin{abstract}
Nuclear collisions at intermediate, relativistic, and ultra-relativistic energies offer 
unique opportunities to study in detail manifold fragmentation and clustering phenomena
in dense nuclear matter. At intermediate energies, the well known processes of nuclear 
multifragmentation -- the disintegration of bulk nuclear matter in clusters of a wide 
range of sizes and masses -- allow the study of the critical point of the equation of 
state of nuclear matter. At very high energies, ultra-relativistic heavy-ion collisions 
offer a glimpse at the substructure of hadronic matter by crossing the phase boundary 
to the quark-gluon plasma. The hadronization of the quark-gluon plasma created in the
fireball of a ultra-relativistic heavy-ion collision can be considered, again, as 
a clustering process. We will present two models which allow the simulation of nuclear 
multifragmentation and the hadronization via the formation of clusters in an interacting 
gas of quarks, and will discuss the importance of clustering to our understanding of 
hadronization in ultra-relativistic heavy-ion collisions.
\end{abstract}

\vfill
\noindent
While most experimental studies concerning clustering and fragmentation of 
matter focus on the scale of atoms and molecules, there are prominent examples of these
phenomena on the more fundamental scale of nuclear matter. In this note, we want to
briefly present two of them: the multifragmentation transition for heated, diluted nuclear
matter, and the clustering of quarks and hadrons at the transition from a quark-gluon-plasma
to a gas of hadrons. The theoretical models we will use to study the relevant physics are the 
Quantum Molecular Dynamics (QMD) for nuclear matter, and the quark Molecular Dynamics (qMD) 
for the subnuclear degrees of freedom, respectively. We will further discuss how clustering
helps to understand data from ultra-relativistic heavy-ion collisions at the 
Relativistic Heavy Ion Collider (RHIC), on the level of clustering both of partons and of hadrons.

\section{Multifragmentation in Nuclear Matter}

\begin{figure}[t]
\parbox[b]{0.35\linewidth}{%
\caption{\label{fig:NuclearClustering}%
\small Results of a Quantum Molecular Dynamics simulation of bulk nuclear matter at different densities.
While at normal nuclear densities (lower row), nuclear matter is distributed homogeneously, 
prominent clustering builds up at diluted densities, $\rho \sim 0.1 \rho_0$ (upper row).}} 
\parbox[b]{0.65\linewidth}{%
\includegraphics[width=\linewidth]{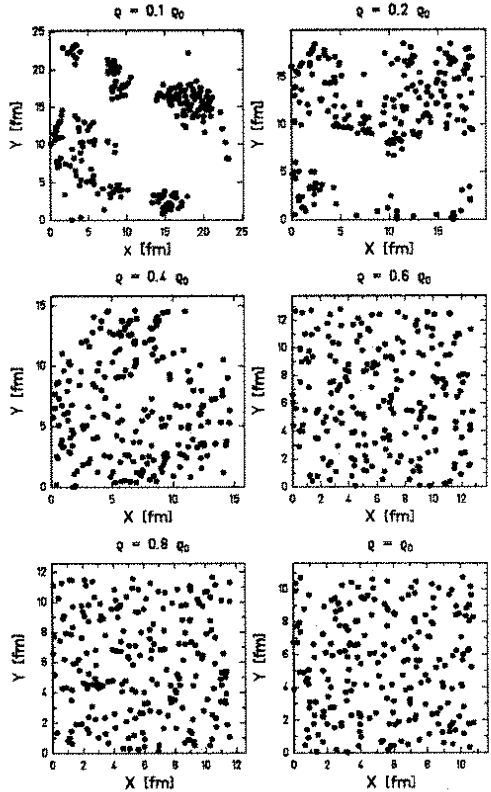}}
\end{figure}

At the heart of all matter, nearly all the mass of every atom is concentrated in the tiny atomic
nucleus, taking roughly $1/10^{15}$ of the volume of the atom. The atomic nucleus is build up of
protons and neutrons, which are bound together by nuclear forces, effective remnants of the fundamental 
strong interaction between quarks and gluons. Understanding the the nuclear forces is essential in order
to understand, for example, which nuclei can be stable, and to gain a complete overview of the chart of isotopes.
From the theoreticians point of view, a possible way to study nuclear forces is to incorporate them
in a model which is then solved numerically on a computer. Such a model is, e.\ g.\, the
Quantum Molecular Dynamics (QMD) of nuclear matter. Here, nucleons are modelled as Gaussian wavepackets, 
with realistic potential interactions corresponding to the nuclear forces, and Fermi statistics is mimicked 
by a Pauli potential~\cite{Peilert:kr}. Results of QMD calculations of bulk nuclear matter at different nuclear 
densities are shown in figure~\ref{fig:NuclearClustering}. While at normal nuclear densities, bulk nuclear 
matter is homogeneous, a strong clustering is observed at low densities ($\rho \approx 0.1\rho_0$). 

\begin{figure}[t]
\parbox[b]{0.45\linewidth}{%
\caption{\label{fig:EoSNuclearClustering}%
\small The Equation of State (EoS) of nuclear matter at different temperatures. Solid lines show
the energy per nucleon for infinite, homogeneous nuclear matter at different temperatures. 
This energy will be lowered significantly if the clustering of nucleons is taken into account (dotted marks).
}}
\parbox[b]{0.52\linewidth}{%
\includegraphics[width=\linewidth]{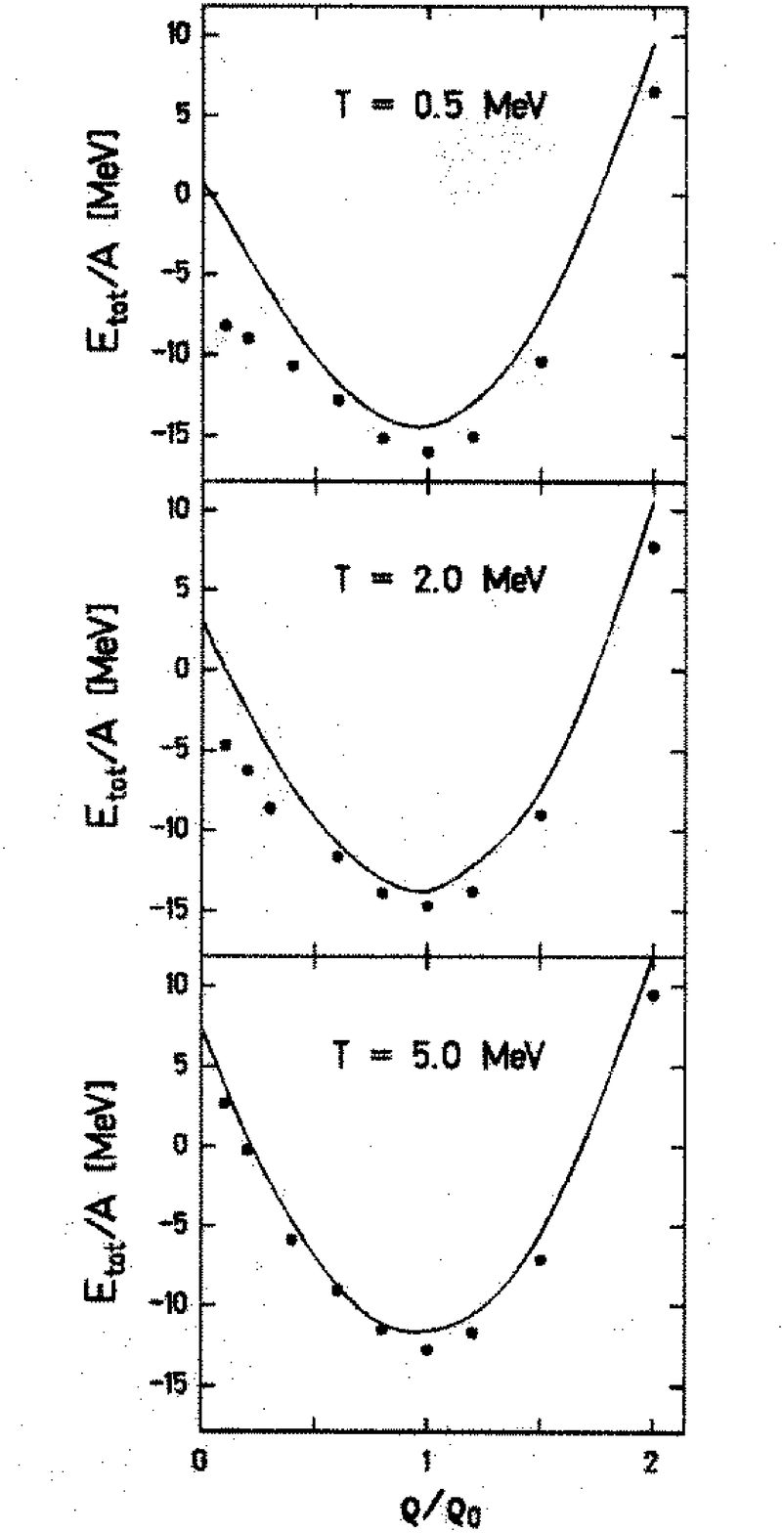}}
\end{figure}

What are the physical consequences of this clustering? While it is difficult to access nuclear forces 
directly by experiment, a lot of information can be gained by the study of the Equation of State (EoS) 
of nuclear matter, which gives the energy per nucleon as a function of nuclear density. The EoS can be 
probed, for example, in nuclear collisions. In a first approximation, looking at homogeneous, infinite nuclear 
matter, the energy per nucleon depends on bulk nuclear density and temperature. These relations are is plotted 
for different temperatures in figure~\ref{fig:EoSNuclearClustering} as solid lines. However, calculations 
with QMD show that allowing for clustering will lower the energy per nucleon. These energy shifts are most 
prominent at low densities and temperatures, where clustering is strongest.

\begin{figure}[t]
\parbox[b]{0.45\linewidth}{%
\caption{\label{fig:Guggenheim}%
\small The Guggenheim plot for finite nuclear matter in different nuclear collisions: Nuclear fragments
populate the the low density (vapour) branch of the coexistence curve of finite nuclear matter.
(from Elliott {\it et al.} \protect\cite{Elliott:2002dk}.)
}}
\parbox[b]{0.52\linewidth}{%
\includegraphics[width=\linewidth]{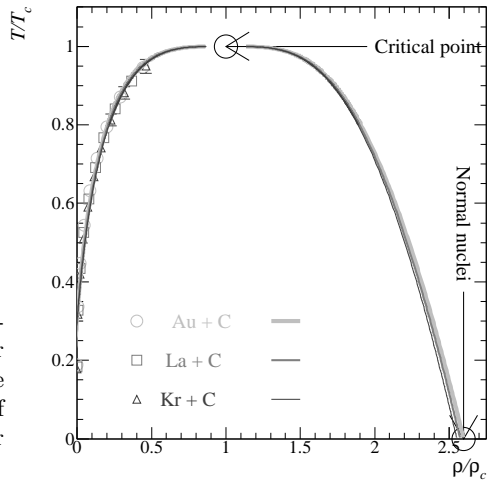}}
\end{figure}

How can these calculations be checked by experiment in the laboratory? On possibility is the analysis
of nuclear collisions at intermediate energies (about 100--500 MeV$/N$). Such collisions yield in a first
stage compressed nuclear matter, which subsequently expands, thereby running through a stage of diluted 
nuclear matter which fragments in clusters of different sizes. Of course, the systems studied in such
collisions are far from representing infinite nuclear matter which exists only in neutron stars, so it is
essential to take into account finite size effects~\cite{KleineBerkenbusch:2001kq}.
It turns out of the study of the cluster size distribution that the fragmentation of 
nuclear matter in these collisions can be understood in terms of a liquid-gas phase transition: 
diluted and heated nuclear matter fragments and evaporates like a Van der Waals fluid! Figure~\ref{fig:Guggenheim} 
shows the corresponding Guggenheim plot, representing the results of this fragmentation analysis~\cite{Elliott:2002dk}.

\section{Clustering and the transition to the quark-gluon plasma}

At the liquid-gas transition of nuclear matter, the substructure of the nucleons does not matter. 
However, as is well known since the 1970s, all hadrons such as protons and neutrons do have such a substructure:
they are composed of quarks and gluons. Hadrons consisting of three quarks are called baryons, hadrons
made up of a quark and an antiquark are called mesons. (Very recently, a short living state consisting of 
five quarks -- a so called pentaquark state, with 4 quarks and 1 anti-$s$-quark and the electric charge of 
the proton -- has been found in nuclear reactions~\cite{Nakano:2003qx}, but we will not discuss this topic further.) 
One may ask whether it is possible to separate single quarks from nucleons by suitable scattering experiments. 
This, however, can not happen, which is a consequence of Quantum Chromodynamics (QCD), the gauge theory describing the
interaction between quarks. 

In QCD, quarks carry a so-called colour charge, which comes in three types (red, green, blue), 
corresponding to the fundamental representation of the gauge group $SU(3)$. This colour charge should 
not be confused with the quark flavour, which can be up, down, strange, charm, top, and bottom, 
where only the first four are relevant in current nuclear collision experiments. The gauge bosons 
mediating the interactions between quarks are called gluons. Since the gauge group $SU(3)$ is non-abelian, 
gluons also interact among themselves. As a consequence, the colour field created by two quarks 
of opposing colour does not spread over all space as in electrodynamics, but is confined to a so-called 
flux tube. This means that the interaction energy between two quarks increases linearly with distance.
A large enough increase of the distance between two quarks hence deposits enough energy in the
flux tube that a new quark-antiquark pair will be created in the flux tube, not allowing a single
quark to escape. For the same reason, all hadrons are colour neutral, hence internally carrying colour
and anticolour (mesons) or three different colours (baryons). 
This property of QCD is called colour confinement.

\subsection{Experimental studies of the quark-gluon plasma}

\begin{figure}[t]
\parbox[b]{0.38\linewidth}{%
\caption{\label{fig:PhaseDiagram}%
\small A schematic view of the phase diagram of nuclear matter. For diluted, cool systems, there is the liquid-gas transition. 
For hot or dense systems, hadronic matter undergoes the transition to the quark-gluon plasma.
}}
\parbox[b]{0.60\linewidth}{%
\includegraphics[width=\linewidth]{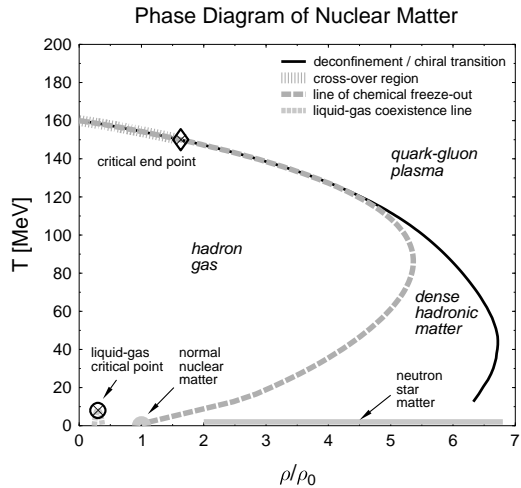}}
\end{figure}

Colour confinement does not mean, however, that quarks must always be bound to hadrons. 
It means that there can be no single, free colour charges. Larger chunks of nuclear matter 
consisting of hadrons can indeed undergo transition to a dense system of free quarks and 
gluons -- this is the transition to the quark-gluon plasma (QGP). Figure \ref{fig:PhaseDiagram} 
shows a simplified, schematic version the corresponding phase diagram of nuclear matter.
Normal nuclear matter from atomic nuclei is at $T=0$~MeV and at $\rho_0 = $112~MeV/fm$^3$. At lower densities 
and slightly higher temperatures, there is the liquid-gas transition with its critical point which manifests 
itself in the multifragmentation of nuclear matter. At zero temperature and higher densities, there is the 
bulk nuclear matter which is found in neutron stars. At higher temperatures there is the deconfinement transition, 
above which quarks and gluons can move freely in the hot and dense system. At the transition to the 
quark-gluon plasma, quark masses drop to their current masses and chiral symmetry is restored, 
which is why the QGP transition is also called chiral transition.

Probably the only place in nature where the quark-gluon plasma transition has ever occurred is the early universe. 
Nevertheless, it is possible to study this transition in the laboratory -- this is the scientific aim of the
ultra-relativistic heavy ion programs at GSI in Darmstadt, CERN, and the RHIC at BNL. In these experiments,
heavy nuclei such as Au or Pb are brought to collisions at energies of $\sqrt{s}_{NN} \approx 7-18$~GeV (CERN-SPS)
or even $\sqrt{s}_{NN} \approx 130-200$~GeV (BNL-RHIC). In such a collision, nuclear matter is compressed, 
and a fireball -- a zone of very hot and dense nuclear matter -- is created, where the transition to 
the QGP state occurs. In the subsequent expansion and cooling of the fireball, the quark-gluon
matter condenses again to a dense, interacting system of hadrons, which further expands and undergoes the chemical 
freeze-out after which there are no more changes in the composition of the system. The final state hadrons are
the particles that can be measured in detectors. Temperatures and chemical potentials which can be extracted
from the measured hadrons at different experiments yield the curve of chemical freeze-out shown in 
figure \ref{fig:PhaseDiagram}.

\begin{figure}[t]
\parbox[b]{0.38\linewidth}{%
\caption{\label{fig:QuarkEvolution}%
\small Time evolution of the number of quarks and anti-quarks in a Pb+Pb 
collision at SPS energies ($\sqrt{s}_{NN} = 17.3$~GeV), as calculated from qMD.
Quarks form clusters of three quarks or of a quark and an antiquark, which are
mapped to baryons and mesons, respectively.
}}
\parbox[b]{0.60\linewidth}{%
\includegraphics[width=\linewidth]{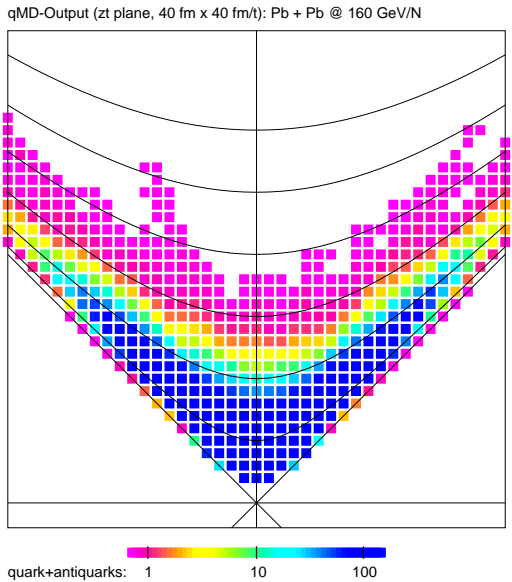}}
\end{figure}

\subsection{Modelling the quark-gluon plasma: qMD}

Since QCD is a very complex theory which is not yet solved analytically, theoretical studies of
the quark-gluon plasma always involve the construction of models. One such model which can be used
to examine the hadronization of an expanding quark-gluon plasma is the quark molecular dynamics (qMD).

The idea of this model is to treat quarks (and antiquarks) as classical particles carrying a colour charge and
interacting via a potential which increases linearly with distance and thus mimics the confining 
properties of colour flux tubes. The relative strength of the coupling depends on the colours of the
quarks involved, and it can be both attractive and repulsive. Thus, the Hamiltonian of the model reads

\begin{equation}
{\mathcal{H}} = \sum_{i=1}^N\sqrt{\vec{\mathbf{p}}_i^2+m_i^2}
+\frac{1}{2}\sum_{i,j}C_{ij}V(\left\vert\vec{\mathbf{r}}_i-\vec{\mathbf{r}}_j\right\vert)\:,
\quad V(r) = -\frac{3}{4}\frac{\alpha_s}{r}+\kappa\,r
\label{eq:Hamiltonian}
\end{equation}

The time evolution of a system of quarks described by this Hamiltonian yields the formation
of clusters of two quarks (quark and antiquark with colour and anticolour) and of three quarks (or three
antiquarks) of three different colours. This is due to the colour-dependency of the interaction, which
favours a redistribution of a homogeneous system in colour neutral clusters. In qMD, these clusters
are mapped on hadronic states according to their masses and quantum numbers such as spin and isospin.

\begin{figure}[t]
\parbox[b]{0.48\linewidth}{%
\includegraphics[width=\linewidth]{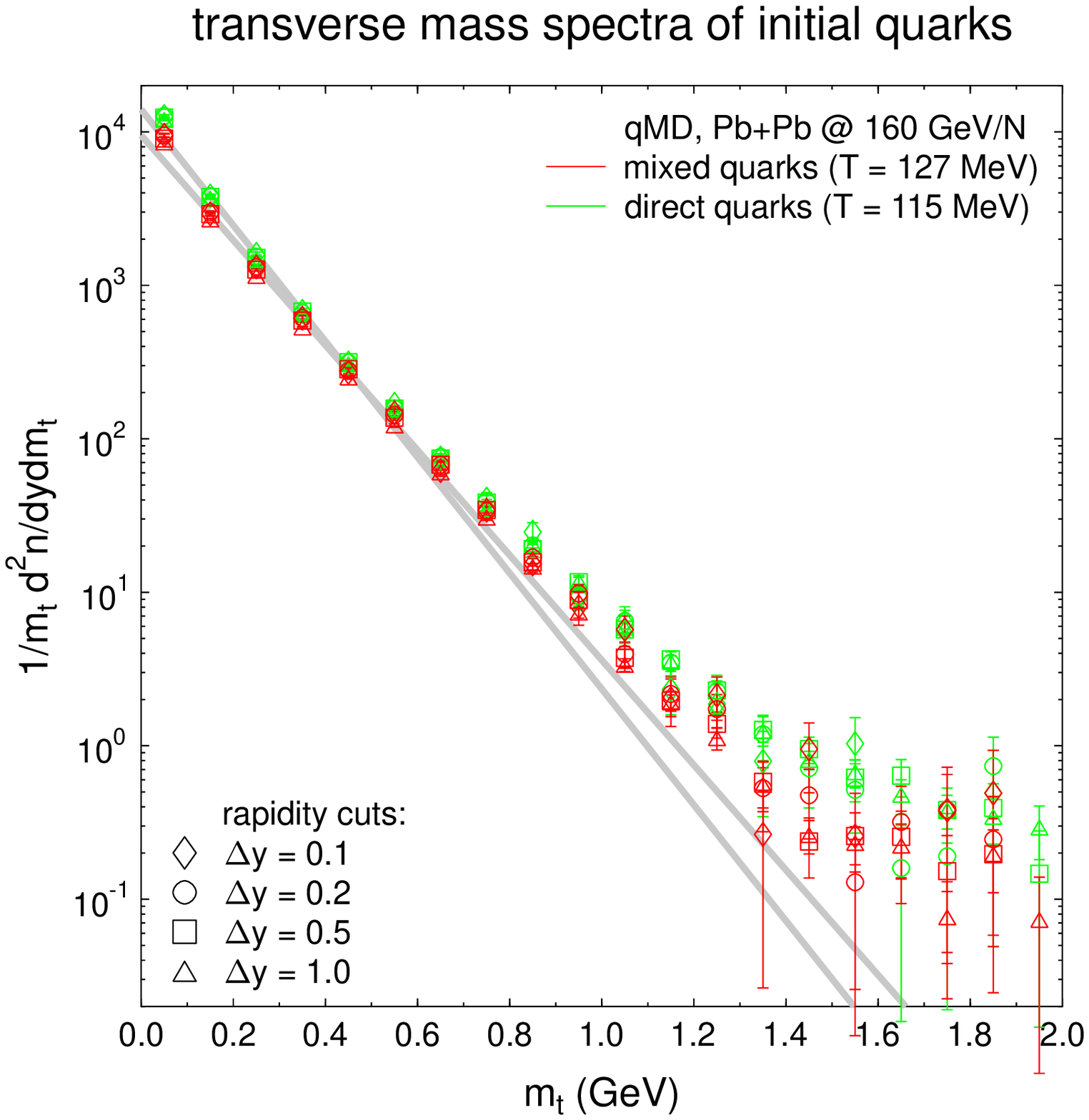}}
\parbox[b]{0.48\linewidth}{%
\includegraphics[width=\linewidth]{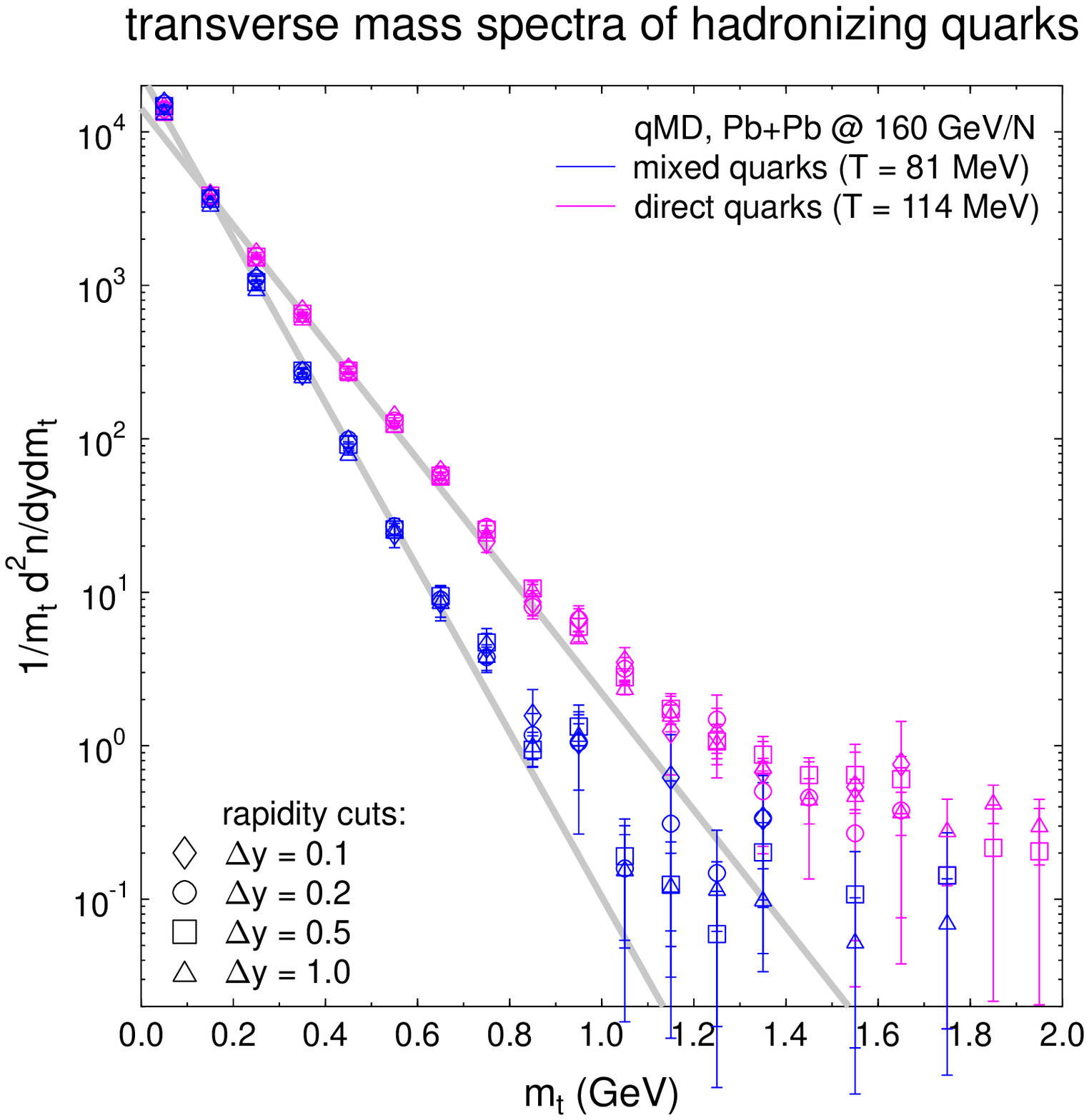}}
\caption{\label{fig:InitialFinalQuarks}%
\small Transverse mass spectra of quarks in the initial phase of the the qMD calculation (left) and at
hadronization (right). The spectra can be fitted quite reasonably with a thermal model. Temperatures
from the slope of the spectra show cooling during the time evolution, which is much stronger for
mixed quarks than for direct quarks.
}
\end{figure}

Starting from this Hamiltonian, Monte Carlo calculations show a transition between two very distinct
phases~\cite{Hofmann:1999jx}, from one dominated by clusters at low temperatures, to a phase of free
quarks at high energies. This can be seen as a simple model of the quark-gluon plasma transition.

\begin{figure}[t]
\parbox[b]{0.38\linewidth}{%
\caption{\label{fig:PartonClusteringFragmentation}%
\small Transverse momentum spectrum of charged hadrons in central Au+Au collisions at RHIC 
energies ($\sqrt{s}_{NN} = 200$~GeV). Data from the PHENIX experiment can be understood 
as showing the sum of two contributions from parton clustering and parton fragmentation, 
where parton fragmentation dominates at $p_\perp > 4$~GeV (top). The transition
from clustering to fragmentation is also seen in the ratio of $\mathrm{p}/\pi^+$ (bottom).
Note different $p_\perp$ ranges of this plot and and the plots in figure~\protect\ref{fig:InitialFinalQuarks}.
(From Fries {\it et al.} \protect\cite{Fries:2003vb}.)
}}\hfill 
\parbox[b]{0.55\linewidth}{%
\includegraphics[width=\linewidth]{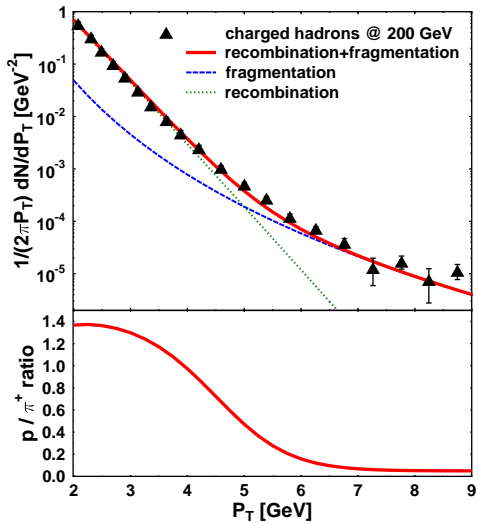}}
\end{figure}

qMD can thus be used to simulate the hadronization of the expanding fire ball in a heavy ion 
collision~\cite{Scherer:2001ap}. Figure~\ref{fig:QuarkEvolution} shows the time evolution of the 
number of quarks and antiquarks in a Pb+Pb collision at SPS ($\sqrt{s}_{NN} = 17.3$~GeV$/N$).
One sees that after an eigenzeit $\approx 15$ fm/c, hadronization is over, what is a very reasonable
result. While qMD allows to calculate experimental observables like particle numbers and momentum spectra,
it offers also the opportunity to look into the microscopic dynamics of hadronization not directly 
accessible to experiment. Figure~\ref{fig:InitialFinalQuarks} shows the transverse momentum spectra 
(a measure of temperature) of ``direct'' quarks (quark correlations from one initial hadron forming 
again the ``same'' hadron) and ``mixed quarks'', which regroup to form new hadronic clusters. While the 
initial temperatures of the two populations of quarks are essentially the same, mixed quarks show a 
much stronger cooling in the expansion than direct quarks. This means that in this model, the
quark system is made up of two distinct subsystems: quarks which interact and interchange their role in 
between the hadronic correlations, and quarks which escape essentially without interaction
from the system, forming hadrons which can be traced back to the initial stage of the collision.

\subsection{Partonic clustering at RHIC}

One of the remarkable discoveries in the experiments at RHIC in Brookhaven has been the suppression 
of pion yields at transverse momenta $p_\perp > 2$~GeV in central Au+Au collisions in comparison 
to p+p collisions. This is generally understood as a consequence of jet quenching and interpreted 
as a strong signal of the creation of a quark-gluon plasma. At the collision energies of RHIC, quarks
and gluons are considered as partons which may scatter with large exchange of momentum. The scattered
partons then fragment into hadrons (as in the string picture mentioned before), which carry away the
transverse momentum. If the scattered partons have to cross a colour-charged medium (as if produced
within a QGP) before fragmentation, they lose energy by processes such as gluon bremsstrahlung, thus
depositing less transverse momentum in the final hadrons. This is the simply physical picture of the
processes yielding jet quenching. 

However, data from RHIC showed a puzzle, which was called the proton/pion anomaly~\cite{Vitev:2001zn}: 
Whereas jet quenching was observed in the transverse momentum spectra of pions as expected, 
it was found to be much smaller for protons and antiprotons. This would mean that the partons 
fragmenting into baryons would suffer less energy loss in the medium than those producing pions, 
which is hard to understand. This missing suppression for baryons is seen best in the ratio 
of protons to pions at transverse momenta of 2--3 GeV, where it surmounts 1 -- a very unusual result. 
A similar riddle showed up in the analysis of elliptic flow~\cite{Lin:2002rw,Voloshin:2002wa}.
These problems can be solved by considering not only parton fragmentation, but also parton
recombination~\cite{Fries:2003vb,Greco:2003xt,Molnar:2003ff}: In the colour-charged medium, scattered
partons can recombine and cluster to form colour-neutral hadrons. This is the same idea as in the qMD 
model. Thus, partons with relatively small transverse momentum can cluster to build up protons with 
$p_\perp \sim$ 2--3 GeV without the need of parton fragmentation. Figure \ref{fig:PartonClusteringFragmentation}
shows how the combination of both parton clustering and fragmentation yields an excellent description 
of the transverse mass spectra of charged hadrons. It also shows how the observed high ratio of $p/\pi$ 
can be understood -- and makes the prediction that this ratio should drop at higher transverse momenta.

It should be noted that the parton recombination involved to explain the RHIC observables do not
include microscopic parton dynamics, which is done in qMD for quarks. Instead, they work by coalescence 
in phase space. It would be tempting to apply qMD to look at these questions for RHIC events. However,
beside the problem of applying the instantaneous potential interaction at RHIC energies, at the
moment it is hard to get enough statistics with qMD simulations to get reliable data for the high $p_\perp$
region which is most interesting. Note that the $m_\perp$ spectra in figure \ref{fig:InitialFinalQuarks} (which
are essentially $p_\perp$ spectra for massless quarks) end in statistical noise at $m_\perp = 2$~GeV, 
which is the lower $p_\perp$ offset of figure \ref{fig:PartonClusteringFragmentation}.

\subsection{Nucleonic clustering at RHIC: Antimatter}

Once partons in a heavy ion collision have fragmented or recombined to hadrons, this is not the 
end of the story as far as clustering phenomena are concerned. The dense hadronic medium in the 
expanding fireball after the transition from the QGP to hadrons allows for many rescattering 
processes. During rescattering, the formation of nuclei and even anti-nuclei by coalescence of
nucleons is possible~\cite{Spieles:1995fs}. 

The STAR collaboration has been looking for anti-deuteron and anti-helium in the final
particle yields of Au+Au collisions at RHIC~\cite{Adler:2001uy}. Production rates of $\bar{\mathrm{d}}$ and 
${}^3\overline{\mathrm{He}}$ were found which are larger than in nucleus-nucleus collisions with 
lower energies -- this can be understood by the much more copiously produced anti-nucleons in 
the nearly net-baryon free fireball of a RHIC event as compared to the net-baryon rich events 
at lower energies. In fact, the production of light anti-nuclei fits very well the expectations
from anti-nucleon coalescence models: anti-nucleons cluster together to form anti-deuteron and anti-helium.

\section{Conclusion}

We have presented two examples of fragmentation and clustering phenomena from nuclear physics
at mediate and high energies. There are many other cases in nuclear physics where clustering and 
fragmentation are important -- strange nuclear matter with such objects as the pentaquark states, 
strangelets and MEMOs are among them. The examples presented here can only give a scarce impression 
of this very rich and interesting field.

\section*{Acknowledgements}

The authors thank Steffen Bass and Marcus Bleicher for helpful hints and fruitful discussions.

\end{document}